\documentstyle[12pt,epsf]{article}
\def\lsim{\mathrel{\vcenter{\hbox{$<$}\nointerlineskip\hbox{$\sim$}}}}

\newcommand{\barr}{\begin{array}}
\newcommand{\earr}{\end{array}}
\newcommand{\beq}{\begin{equation}}
\newcommand{\eeq}{\end{equation}}
\newcommand{\bea}{\begin{eqnarray}}
\newcommand{\eea}{\end{eqnarray}}

\newcommand{\prlr}[3]{Phys.\ Rev.\ Lett.\ {\bf #1},~#2~(#3)}

\newcommand{\prdr}[3]{Phys.\ Rev.\ {\bf #1}, #2~(#3)}
\newcommand{\plbr}[3]{Phys.\ Lett.\ {\bf #1},~#2~(#3)}
\newcommand{\npbr}[3]{Nucl.\ Phys.\ {\bf #1},~#2~(#3)}

\newcommand{\zp}[3]{Z.\ Phys.\ {\bf #1}, #2 (#3)}

\begin{document}
\begin{titlepage}
\rightline{\vbox{\halign{&#\hfil\cr
&CUMQ/HEP 116\cr
&HIP-2001-18/TH\cr
&\today \cr}}}
\vspace{0.5in}
\begin{center}

{\Huge Spontaneous R-Parity violation bounds}
\\
\medskip
\vskip0.4in
\normalsize{M.~Frank$^{a}$\footnote[1] {e-mail
mfrank@vax2.concordia.ca} and K. Huitu$^{b}$\footnote[2]{e-mail
huitu@pcu.helsinki.fi} }
\smallskip
\medskip

{\sl $^{a}$Department of Physics, Concordia University, 1455 De Maisonneuve
Blvd. W.\\ Montreal, Quebec, Canada, H3G 1M8}

{\sl $^{b}$Helsinki Institute of Physics, P.O.Box 64,
FIN-00014 University of Helsinki, Finland }
\\
\end{center}
\vskip1.0in

\begin{center}
{\large\bf Abstract}
\end{center}

\smallskip

We investigate bounds from tree-level and one-loop processes
in generic supersymmetric models
with spontaneous R-parity breaking in
the superpotential. We analyse the bounds from a general point of
view.
The bounds are applicable both for all models with spontaneous
R-parity violation and for explicit bilinear R-parity violation based
on general lepton-chargino
     and neutrino-neutralino mixings. We find constraints from
semileptonic B, D and K decays, leptonic
decays of the
$\mu$ and $\tau$, electric dipole moments, as well as
bounds for the anomalous magnetic moment of the
muon.

\smallskip

PACS number(s): 12.60.Jv, 11.30.Fs, 14.80.Ly, 13.10.+q

\end{titlepage}
\baselineskip=20pt
\newpage
\pagenumbering{arabic}

\section{\bf Introduction}

While supersymmetry appears to be the best scenario for physics
beyond the Standard Model (SM), most of the
early studies have been made in the context of the minimal
supersymmetric standard model (MSSM), the supersymmetric
analog of the Standard Model. This model assumes conservation of a
discrete symmetry called the R-parity, which
is related to baryon number, lepton number and spin through
$R=(-1)^{(3B+L+2S)}$. Under this symmetry all the
Standard Model particles are R-even, while their superpartners are
R-odd. With this assumption the
supersymmetric particles must be pair produced, every supersymmetric
particle decays into another and the
lightest of them is stable. Although the MSSM has some attractive
features, neither gauge invariance nor
supersymmetry require R-conservation, so within the MSSM
R-conservation is imposed as a symmetry of the model.
The most general supersymmetric extension of the Standard Model
containing explicit R-violating interactions has
received a lot of attention lately as one of the possible
explanations for neutrino masses and oscillations.
Numerous detailed analyses of the explicit (bilinear or trilinear)
Yukawa couplings have appeared in the
literature along with constraints on these couplings
\cite{trilinear, bilinear}.

Less attention has been given to the possibility that R is an exact
symmetry of the Lagrangian, but broken
spontaneously through the Higgs mechanism. This would occur through
scalar neutrinos acquiring non-zero
vacuum expectation values:
\bea
\langle\tilde\nu_{L_i}\rangle \ne 0~~~;~~~
\langle\tilde\nu_{R_i}\rangle \ne 0
\eea
Such a breaking is natural in scenarios beyond the MSSM, such as for
instance in the left-right supersymmetric
model (LRSUSY) where spontaneous R-parity breaking avoids the existence of a
charge-violating minimum in
the superpotential \cite{LRR}. If spontaneous R-parity violation occurs in
the  absence of any additional gauge symmetries,
it will lead to the existence of a physical massless Nambu-Goldstone
boson, called a Majoron \cite{majoron}.
Phenomenological studies of spontaneous R-parity breaking have
mostly concentrated on the experimental
consequences of the Majoron, the most significant of which is the
increase of the invisible $Z^0$ width by an
amount equivalent to 1/2 of a light neutrino family \cite{majZ}.

Another phenomenologically interesting consequence of the
spontaneous R-symmetry breaking is that it
     introduces new terms in the Lagrangian, as compared to the
Yukawa couplings of the explicit trilinear
R-parity breaking. In particular it introduces interactions with
gauginos
which have the feature that they are
sfermion-mass independent.
Precision measurements of rare processes
put rather strong constraints on such
spontaneous R-breaking terms.
These interactions appear in explicit bilinear breaking as soft
dimensionful Higgs-lepton superfield mixing parameter.

In this work we study the phenomenological implications of
spontaneous R-parity  breaking in the supersymmetric
Lagrangian. We analyse the general form of the
gaugino-higgsino-lepton mixing and set the most general bounds
on the mixing elements based on rare tree-level and one-loop level
processes. The advantage of setting bounds
on the mixing elements lies in their generality: they apply to
any supersymmetric model with spontaneous R-parity breaking,
or even to a SUSYGUT scenario with
an enriched gauge sector.  Our paper is
organized as follows: we describe and parametrize spontaneous
R-parity breaking in section 2, then discuss tree-level
constraints in section 3, one-loop level constraints in section 4,
before reaching our conclusion in section 5.

\section{\bf Spontaneous R-parity breaking}

Specific superpotentials can be designed to violate R-parity and
lepton number spontaneously \cite{r}. We
concentrate here on the phenomenological consequences. As a
consequence of spontaneous R-parity breaking, the
sneutrino fields
${\tilde
\nu}_i$ acquire non-zero vacuum expectation values
$\langle{\tilde \nu}_i\rangle\ne 0$.
In order to have the spontaneous breaking of the R-parity, new fields
have to be added to the MSSM spectrum. In order to set spontanous R
parity breaking
in perspective, we outline briefly the main
features of two models present in the literature.

In the model proposed in \cite{rc}, a superpotential which conserves
total lepton number and R-parity is constructed. Additional
fields $(\Phi, \nu^c_i, S_i)$ are introduced, which are singlets
under $SU(2)_L \times
U(1)_Y$ and which carry a conserved lepton number assigned as
$(0, -1, 1)$. The superpotential has the form:
\bea
W=h_uQH_uU+h-dQH_dD+(h_0H_uH_d-\epsilon^2)\Phi+h_{\nu}LH_u\nu^c+h\Phi
S\nu^c.
\eea
R-parity is broken spontaneously, if one or more of the
singlets have a
vev: $v_R=\langle {\tilde \nu}^c_{\tau}\rangle$,
$v_S=\langle S_{\tau}\rangle$ and
$v_L=\langle {\tilde
\nu}_{L
\tau}\rangle$. The vev of the isodoublet Higgs
will drive the electroweak symmetry breaking and allow fermion masses
in the usual fashion.
The bounds on the sneutrino vev's have been considered in the
bilinear R-parity breaking model.
{}From the Superkamiokande data, the constraints for the vev's are
obtained as \cite{BFK}
\bea
&\langle \tilde\nu_e\rangle /\sqrt{M_{SUSY}/100\,{\rm GeV}}&\le 90\, {\rm keV},
\nonumber\\
76\, {\rm keV}\le
&\langle\tilde\nu_{\mu,\tau}\rangle /\sqrt{M_{SUSY}/100\,{\rm GeV}}&
\le 276 \, {\rm keV}.
\eea
In \cite{r} it has been found that $t-b-\tau$ unification is allowed
for $\langle\tilde\nu_\tau\rangle \lsim 5 $ GeV and $b-\tau$
unification is allowed
for $\langle\tilde\nu_\tau\rangle \lsim 50 $ GeV.

In the left-right supersymmetric model, based on the gauge symmetry
$SU(2)_L \times SU(2)_R
\times U(1)_{B-L}$, the R-parity is a discrete subgroup of the
$U(1)_{B-L}$. It is possible that in the
process of spontaneous symmetry breaking this model develops a
minimum which violates R-parity; indeed
in some versions of LRSUSY this breaking is inevitable \cite{LRR,hpp}. In
the minimal version of this model,
the most general gauge invariant superpotential must contain, in
addition to the usual left and
right-handed quark and lepton doublets, two Higgs bidoublets $\Phi_u$
and $\Phi_d$
and four Higgs triplet superfields $\Delta_{L,R}$ and $\delta_{L,R}$.
The superpotential
corresponding to this minimal field content is:
\bea
W_{min}&=&Q^Ti \tau_2(h_{\Phi_u} \Phi _u + h_{\Phi_d} \Phi_d) Q^c +
L^T i \tau_2 (h_{\Phi_u}
\Phi_u + h_{\Phi_d}
\Phi_d)L^c \nonumber \\
&+& h_{\Delta}(L^T i \tau_2 \delta_L L+ L^{cT} i \tau_2
\Delta_RL^c)+\mu_{ij} Tr(i \tau_2 \Phi_i^T i
\tau_2
\Phi_j) + \mu_{\Delta} (\Delta_L \delta_L +\Delta_R \delta_R)
\eea
In the minimal model, breaking parity spontaneously at the
renormalizable level is always accompanied
by spontaneous R parity breaking. This may be cured by adding more
fields to the theory \cite{LRR}. One is
left with a relatively low $SU(2)_R$ breaking scale, with the
spontaneously broken R-parity driven by
$\sigma_R= \langle {\tilde \nu}^c \rangle \ne 0$.

In this work  we will not assume any particular model for the breaking.
Instead we will study  interactions typical for this class of
models.
    In what follows, we will
present the formulas with the MSSM particle content.
This would effectively be the case if the other fields in the
model decouple.
However, the formalism described can be extended straightforwardly to
richer matter/gauge sectors, such as e.g,
MSSM with right-handed neutrinos, where both the left-handed and the
right-handed sneutrinos can acquire a vev, or left-right model, where
the number of gauginos and higgsinos is larger than in the MSSM.

Within the minimal field content, the chargino-lepton mixing matrix becomes
$5 \times 5$ matrix and the neutralino-neutrino matrix a
$7 \times 7$ matrix. The mass eigenstate fields can be written as:
\bea
\label{lhmix}
\Psi_i^0=N_{ij}\Psi^{\prime
0}_j,~~~~~\Psi_i^{+}=V_{ij}\Psi^{\prime
+}_j,~~~~~\Psi_i^{-}=U_{ij}\Psi^{\prime
-}_j
\eea
for, respectively, the neutral and charged fields, where the weak
eigenstates are:
\bea
\Psi^{\prime 0 T}_j&=&( -i\lambda', -i\lambda_3, {\tilde H}^0_1, {\tilde
H}^0_2,\nu_i),~~~i=e,\mu,\tau
\label{neutr}
\\
\Psi^{\prime - T}_j&=&( -i\lambda_-, {\tilde H}^-_1,e_L^-, \mu_L^-,
\tau_L^-),\\
\Psi^{\prime + T}_j&=&( -i\lambda_+, {\tilde H}^+_2, e_L^+, \mu_L^+, \tau_L^+).
\label{charp}
\eea
are the $U(1)_Y$ gauginos, $\lambda_{3,+,-}$ are the $SU(2)_L$
gauginos, and ${\tilde H}$
are the higgsinos.
To  extend to more complicated particle contents one needs to
add more gauginos and higgsinos in Eqs. (\ref{neutr})-(\ref{charp}).

The relevant part of the interaction Lagrangian becomes, for
quarks-squarks-charginos:
\bea
{\cal L}_{q {\tilde q^{\prime}} \Psi^+} &=& -g \sum_i \left \{{\bar \Psi}_i^+ [
(U^{\ast}_{i1}P_L-\frac{m_{u_k}V_{i2}}{\sqrt{2}M_W \sin \beta}P_R)u_k
{\tilde d}_{Lk}^{\ast} \right. \nonumber
\\
     &-& \frac{m_{d_k}U_{i2}^{\ast}}{\sqrt{2}M_W \cos \beta}P_R u_k
{\tilde d}_{Rk}^{\ast} ]+ {\bar
u}_k(U_{i1}P_R-\frac{m_{u_k}V_{i2}^{\ast}}{\sqrt{2}M_W \sin
\beta}P_L)\Psi_i^+ {\tilde d}_{Lk} \nonumber
\\
     &-& \frac{m_{d_k}U_{i2}}{\sqrt{2}M_W \cos \beta}P_L\Psi_i^+ {\tilde
d}_{Rk}]+ {\bar
\Psi}_i^{+ c}(V_{i1}^{\ast}P_L-\frac{m_{d_k}U_{i2}}{\sqrt{2}M_W \cos
\beta}P_R)d_k {\tilde u}_{Lk}^{\ast}
     \nonumber
\\
     &-& \frac{m_{u_k}V_{i2}^{\ast}}{\sqrt{2}M_W \sin \beta}P_R d_k
{\tilde u}_{Rk}^{\ast} ]+ {\bar
d}_k(V_{i1}P_R-\frac{m_{d_k}U_{i2}^{\ast}}{\sqrt{2}M_W \cos
\beta}P_L)\Psi_i^{+ c} {\tilde u}_{Lk}
     \nonumber \\
     &-& \left. \frac{m_{u_k}V_{i2}}{\sqrt{2}M_W \sin \beta}P_L\Psi_i^{+
c} {\tilde u}_{Rk}] \right\}
\label{q_sq_c}
\eea
and for quarks-squark-neutralinos:
\bea
{\cal L}_{q {\tilde q} \Psi^0} &=&-\sqrt{2} \sum_j \left\{ {\bar u}_k
\{[ee_uN_{j1}+\frac{g}{\cos
\theta_W}(1/2-e_u \sin^2 \theta_W N_{j2})]P_R \Psi_j^0 {\tilde
u}_{Lk} \right. \nonumber \\
&+& \frac{g m_{u_k}}{2 M_W \sin \beta} N_{j4}^{\ast}P_L \Psi_j^0
{\tilde u}_{Lk} -
[ee_uN_{j1}^{\ast}- (\frac{g e_u \sin^2 \theta_W}{\cos
\theta_W}) N_{j2}^{\ast})]P_L \Psi_j^0 {\tilde u}_{Rk} \nonumber \\
&+& \frac{g m_{u_k}}{2 M_W \sin \beta} N_{j4} P_R \Psi_j^0 {\tilde
u}_{Rk} \} \nonumber \\
&+&{\bar d}_k \{[ee_d N_{j1}- \frac{g}{\cos
\theta_W}(1/2+ e_d \sin^2 \theta_W N_{j2})]P_R \Psi_j^0 {\tilde
d}_{Lk} \nonumber \\
&+& \frac{g m_{d_k}}{2 M_W \cos \beta} N_{j3}^{\ast}P_L \Psi_j^0
{\tilde d}_{Lk} -
[ee_d N_{j1}^{\ast}- (\frac{g e_u \sin^2 \theta_W}{\cos
\theta_W}) N_{j2}^{\ast})]P_L \Psi_j^0 {\tilde d}_{Rk} \nonumber \\
&+& \left. \frac{g m_{d_k}}{2 M_W \cos \beta} N_{j3} P_R \Psi_j^0
{\tilde d}_{Rk} \} \right \}.
\label{q_sq_n}
\eea
Similar expressions are obtained for the lepton-slepton interactions.
Note
that in both parts of the Lagrangian, Eqs. (\ref{q_sq_c}) and (\ref{q_sq_n}),
we get interactions which do not depend on mass, but depend on
the mixing with the gaugino.
This is a major difference compared to the explicit trilinear
R-parity breaking,
in which only those terms which depend on mass and also on the mixing
with higgsino are present.

Assuming that the new fields, which transform the potential so that
the R-parity breaks, do not mix with the MSSM higgsinos and gauginos,
the mass matrices for the spontaneous R-parity breaking and
explicit bilinear R-parity breaking are very
similar.
With this assumption,
the mass matrices become ($i,j=e,\mu,\tau$):
\bea
(M)_{\Psi^{\pm}}&=&
\left(
\begin{array}{ccc}
     M_2& \frac{1}{\sqrt{2}}gv_u & 0 \\
     \frac{1}{\sqrt{2}}gv_d &\mu  &\frac{1}{\sqrt{2}}h_i\langle 
{\tilde \nu}_{Li}\rangle \\
\frac{1}{\sqrt{2}}g \langle {\tilde \nu}_{Lj}\rangle & 
h_{\nu_{ij}}\langle \tilde\nu_{Rj}\rangle &
\frac{1}{\sqrt{2}}h_iv_d \delta_{ij} ,
\end{array}
\right)
\eea
for the chargino-lepton (where $h_i$ are Yukawa couplings from the
R-conserving Lagrangian
$h_iL_iH_1E_i$), and:
\bea
(M)_{\Psi^0}&=&
\left(
\begin{array}{ccccc}
     M_1& 0 &\frac{1}{\sqrt{2}}g'v_d &\frac{1}{\sqrt{2}}g'v_u &
\frac{1}{\sqrt{2}}g'\langle {\tilde
\nu}_{Li}\rangle
\\ 0 & M_2 & \frac{1}{\sqrt{2}}gv_d &\frac{1}{\sqrt{2}}gv_u
&\frac{1}{\sqrt{2}}g'\langle {\tilde
\nu}_{Li}\rangle \\ \frac{1}{\sqrt{2}}g'v_{d}& 
\frac{1}{\sqrt{2}}gv_{d}& 0 & -\mu & 0\\
     \frac{1}{\sqrt{2}}g'v_{u}&\frac {1}{\sqrt{2}}gv_{u}&  -\mu & 0&
-h_{\nu_{ij}}\langle \tilde\nu_{Rj}\rangle \\
     \frac{1}{\sqrt{2}}g'\langle {\tilde \nu}_{Li}\rangle 
&\frac{1}{\sqrt{2}}g\langle {\tilde
\nu}_{Li}\rangle &  0& -h_{\nu_{ij}}\langle \tilde\nu_{Rj}\rangle &
0
\end{array}
\right)
\eea
for the neutralino-neutrino.
The lightest mass eigenstates obtained by
diagonalizing the mass matrices correspond to the neutrinos and
charged leptons. By
rotating the MSSM  Lagrangian to the new mass eigenstates one obtains
new lepton-flavor
violating interactions. The mixing matrices induced by the ${\cal
L}_{LH}$ are listed below.
     For the charged sector the matrices are \cite{af}:
\bea
\label{chargmassmatr}
(U)^{\ast}&=&
\left(
\begin{array}{cc}
     U_R(1-\frac{1}{2} \xi^{L^T}\xi^{L \ast})& -V_L
\xi^{L \ast} \\
     U_R \xi^{L^T} &V_L(1-\frac{1}{2}\xi^{L \ast} \xi^{L^T})
\end{array}
\right), \\
(V)^{\dagger}&=&
\left(
\begin{array}{cc}
(1-\frac{1}{2} \xi^{R^T}\xi^{R \ast})U_L^{\dagger}&  \xi^{R
\ast}U_L^{\dagger} \\
     - \xi^{R^T} V_R^{\dagger} & (1-\frac{1}{2}\xi^{R \ast}
\xi^{R^T})V_R^{\dagger}
\end{array}
\right).
\eea
     In the neutral fermion sector the mixing matrix is:
\bea
\label{neutrmassmatr}
(N)^{\ast}&=&
\left(
\begin{array}{cc}
N^{0\ast}(1-\frac{1}{2} \xi^{\dagger}\xi)&
-V^{(\nu)^{T}} \xi \\
     N^{0\ast} \xi^{\dagger} &V^{(\nu)^{T}}(1-\frac{1}{2}\xi \xi^{\dagger})  ,
\end{array}
\right)
\eea
In the above equations the parameters $\xi^L,~\xi^R,~\xi$  represent
mixing beween the MSSM sector
     matrices (corresponding to the matrices $U_L$, $U_R$ for chargino
and $N^0$ for the neutralino in MSSM) and the
lepton ($V_L$,
$V_R$) or neutrinos ($V^{(\nu)}$) mixing matrices. The relationship
between these
matrices and the MSSM matrices is:
\bea
U_R M_{\chi^{\pm}}U_L^{\dagger}&=& \mbox{Diag} \{M_{\chi_i^{\pm}}\} ,
\\
V_L M^{(l)} V^{\ast}_R &=& \mbox{Diag}\{m_{l_i}\},\\
N^{0\ast}M_{\chi^0}N^{o\dagger} &=& \mbox{Diag}\{M_{\chi_i^0}\}, \\
V^{(\nu)^{T}} m_{eff} V^{\nu} &=&
\mbox{Diag}\{m_{\nu_e},m_{\nu_\mu},
m_{\nu_\tau}\},
\eea
where the mixing parameters $\xi^L,~\xi^R,~\xi$ are, for $i=1,2,3$:
\bea
\label{lcmix}
\xi_{i1}^{\ast L}&=&\frac {g_2}{\sqrt{2} \det M_{\chi^{\pm}}}\Lambda_i ,\\
\xi_{i2}^{\ast L}&=& \frac { h_{\nu_{ij}}\langle \tilde\nu_{Rj}\rangle }{\mu}
-\frac {g_2 \sin \beta
M_W}{\mu \det
M_{\chi^{\pm}}}\Lambda_i ,\\
\xi^{\ast R}&=&M^{(l)\dagger}\xi^{\ast L}(M_{\chi^{\pm}}^{-1})^T,\\
\xi_{i1}&=&\frac {g_1 M_2 \mu}{2 \det M_{\chi^0}}\Lambda_i ,\\
\xi_{i2}&=& -\frac {g_2 M_1 \mu}{2 \det M_{\chi^0}}\Lambda_i ,\\
\xi_{i3}&=& \frac { h_{\nu_{ij}}\langle \tilde\nu_{Rj}\rangle }{\mu}
+\frac {g_2 (M_1 + \tan^2 \theta_W
M_2) \sin \beta M_W}{2
\det M_{\chi^0}}\Lambda_i ,\\
\xi_{i4}&=& -\frac {g_2 (M_1 + \tan^2 \theta_W M_2) \cos \beta M_W}{2 \det
M_{\chi^0}}\Lambda_i.
\eea
Here $\chi^\pm$ and $\chi^0$ denote the MSSM charginos and neutralinos.
In the above expressions $\Lambda_i=\mu\langle {\tilde 
\nu}_{Li}\rangle -\langle H_1\rangle
h_{\nu_{ij}}\langle \tilde\nu_{Rj}\rangle $
     is a measure of the misalignment and is small, but must be
essentially nonzero for
neutrinos to have a mass. Also, in the above $\mu$ is the MSSM bilinear
Higgs coupling, and
$\det M_{{\chi}^{\pm}}=M_2\mu-\sin 2\beta M^2_W$ is the determinant
of the MSSM chargino
mass matrix; $\det M_{\chi^0}= M_W^2\mu
\sin 2\beta(M_1+M_2 \tan^2 \theta_W)-M_1M_2 \mu^2$ is the determinant
of the MSSM
neutralino mass matrix.

     The spontaneous R-parity breaking is driven by $\langle {\tilde
\nu}_{Li}\rangle $ and $\langle {\tilde \nu}_{Ri}\rangle $.
     The mass matrices in bilinear R-parity breaking can be obtained
by making the substitution $h_{\nu ij} v_{Rj}
\rightarrow \epsilon_j$, which makes it easy to read the mixing
matrices from one case to the other.

The parametrization we presented above is only one of the ones
available in the literature.
    Others exists, most notable the single vev parametrization
\cite{singlevev}. In the
next sections we will present bounds on the matrix elements
themselves coming from
phenomenological constraints, which are independent of any
parametrization chosen.

\section{\bf Tree-level bounds on matrix elements from rare decays}
\subsection{\bf Semileptonic B, D and K decays}

In this section we investigate constraints arising from rare decays
of the B, D and K mesons, as well three-body
lepton number violating decays of the $\mu$ and $\tau$, all of which
can occur at tree-level and all of which
put bounds of spontaneous R-parity violating matrix mixing elements.
Since all these decays occur at tree-level
through an exchange of a scalar fermion, we will employ throughout
the notation:
\bea
q_i = \left (\frac{100~GeV}{m_{{\tilde q}_i}^2} \right)^2, \;
l_i = \left (\frac{100~GeV}{m_{{\tilde l}_i}^2} \right)^2, \;
n_i = \left (\frac{100~GeV}{m_{{\tilde \nu}_i}^2} \right)^2
\eea
with $q= u,d$, and $i=1,2,3$ represents the three quark families,
$l_i=e, \mu, \tau$ and $\nu_i=\nu_e,\nu_{\mu},
\nu_{\tau}$.

We first analyse the semileptonic decays of the K, D and B-mesons.
The effective Lagrangians relevant for semileptonic decays of the
B-mesons are:
\bea
{\cal L}_{eff}(b \rightarrow q l_i \nu_j)=-K_{qb}
\frac{4G_F}{\sqrt{2}} [{\cal A}_{ij}^q({\bar
q}\gamma^{\mu}P_Lb)({\bar l}_i\gamma_{\mu}P_L\nu_j)-{\cal B}_{ij}^q({\bar
q}P_Rb)({\bar l}_iP_L\nu_j)],
\eea
and
\bea
L_{eff}(b\rightarrow ql^+_il^-_j) =-K_{qb}\frac{4G_F}{\sqrt{2}}
[{\cal C}_{ij}^q({\bar
q}\gamma^{\mu}P_Lb)({\bar l}_i\gamma_{\mu}P_Ll_j)-{\cal D}_{ij}^q({\bar
q}P_Rb)({\bar l}_iP_Ll_j)],
\eea
where $K$ is the CKM matrix.
The Lagrangian is similar for all the semileptonic decays with the
appropriate substitutions for the $b$
quark. The leptonic branching ratios for the processes $b \rightarrow
e \nu X$ and $b \rightarrow \mu
\nu X$ measured by the L3 Collaboration \cite{L3} are:
\bea
BR(b \rightarrow e \nu X) & = &(10.89 \pm 0.55)\times 10^{-2},\nonumber \\
BR(b \rightarrow \mu \nu X) & = &(10.82 \pm 0.61)\times 10^{-2}.
\eea
These decay processes can occur at tree-level through either
${\tilde d}$ and $\tilde b$ or ${\tilde u}$ and $\tilde t$
exchanges. They set bounds on both the higgsino and the gaugino
couplings in spontaneous R-parity violating
models.

The branching
ratios for semileptonic decays into charged leptons will set bounds
on the chargino-lepton mixing elements only.
The present measurements of the branching ratios of the
$b
\rightarrow sl_j^+l_i^-$ give the following upper bounds (at 90\%
C.L.) \cite{Partdata}
\bea
BR(b \rightarrow se^+e^-) &< & 5.7 \times 10^{-5}, \nonumber \\
BR(b \rightarrow s\mu^+ \mu^-) &< & 5.8 \times 10^{-5}, \nonumber \\
BR(b \rightarrow se^{\pm} \mu^{\mp}) &< & 2.2  \times 10^{-5}.
\eea
     The experimental bounds on the first two are almost one order of
magnitude larger than the SM expectation: the
last process is forbidden in SM because of separate conservation of
each lepton flavor number.
The bounds obtained are listed in Table 1.
They  involve products of neutralino-neutrino and
chargino-lepton mixing matrices.

\vfill

\newpage

\begin{center}
{{\bf Table 1}: Analytic bounds on mixing matrices of chargino-leptons
$U_{ij},~V_{ij}$ (with $j=3,4,5$
corresponding to $e$, $\mu$ and $\tau$) and neutralino-neutrinos
$N_{ij}$ (with $j=5,6,7$
corresponding to $\nu_e,
\nu_{\mu}$ and $\nu_{\tau}$) from B rare semileptonic
decays.}
\vskip0.2in
\begin{tabular}{|l|l|l|c|}
\hline
gaugino type &
higgsino type
& bound & process
\\
\hline
$-\sqrt{2}|K_{td}|\left\{e [V_{31}e_dN_{i1}^{\ast}(d_1+d_3)
\right.$&
$g^2|K_{td}|\left[\frac{(m_um_dd_1N^{\ast}_{i3}U_{32}
+m_bm_td_3N_{i3}U_{32}^{\ast})}
{2M_W^2 \cos^2\beta}\right.$&
$4.8 \times 10^{-3}$&$b\rightarrow u e \nu_i$\\
$-V_{31}^{\ast}e_uN_{i1}(u_1+u_3)]$ &
$\left.+\frac{
(m_um_dd_1N^{\ast}_{i4}V_{32}+m_bm_td_3N_{i4}V_{32}^{\ast})}{2M_W^2 \sin^2
\beta}\right]$
&&\\
$ -g\left[\frac {(1/2+e_d\sin^2
\theta_W)(d_1+d_3)}{\cos \theta_W}V_{31}N_{i2}^{\ast}\right.$&&&\\
$\left.\left.-\frac {(1/2-e_u\sin^2\theta_W)(u_1+u_3)}{\cos \theta_W}
V_{31}^{\ast}N_{i2}\right]\right\}$&&&\\
&&&\\

$-\sqrt{2}|K_{td}|\left\{e [V_{41}e_dN_{i1}^{\ast}(d_1+d_3)\right.$
&
$g^2|K_{td}|\left[
\frac{(m_um_dd_1N^{\ast}_{i3}U_{42}+m_bm_td_3N_{i3}U_{42}^{\ast})}
{2M_W^2 \cos^2 \beta}\right.$
&$5.3 \times 10^{-3}$&$b\rightarrow u \mu \nu_i$\\
$\left.-V_{41}^{\ast}e_uN_{i1}(u_1+u_3)\right]$&$\left.+\frac{
(m_um_dd_1N^{\ast}_{i4}V_{42}+m_bm_td_3N_{i4}V_{42}^{\ast})}{2M_W^2 \sin^2
\beta}\right]$&&\\
$-g\left[\frac {(1/2+e_d\sin^2
\theta_W)}{\cos \theta_W}N_{i2}^{\ast}](d_1+d_3)\right.$&&&\\
$\left.\left.-\frac {(1/2-e_u\sin^2
\theta_W)}{\cos \theta_W}N_{i2}](u_1+u_3)\right]\right\}$&&&\\
&&&\\

$-\sqrt{2}|K_{td}|\left\{e
\left[V_{51}e_dN_{i1}^{\ast}(d_1+d_3)\right.
\right.$&
$g^2|K_{td}|\left[\frac{(m_um_dd_1N^{\ast}_{i3}U_{52}
+m_bm_td_3N_{i3}U_{52}^{\ast})}
{2M_W^2 \cos^2 \beta}\right.$&
$3.1 \times 10^{-3}$&$b\rightarrow u \tau \nu_i$\\
$\left.-V_{51}^{\ast}e_uN_{i1}(u_1+u_3)\right] $&$\left.+\frac{
(m_um_dd_1N^{\ast}_{i4}V_{52}+m_bm_td_3N_{i4}V_{52}^{\ast})}{2M_W^2 \sin^2
\beta}\right]$&&\\
$-g\left[\frac {(1/2+e_d\sin^2
\theta_W)}{\cos \theta_W}N_{i2}^{\ast}(d_1+d_3)\right.$&&&\\
$\left.\left.-\frac {(1/2-e_u\sin^2
\theta_W)}{\cos \theta_W}N_{i2}](u_1+u_3)\right]\right\}$&&&\\
&&&\\
$g^2(u_2+u_3)
V_{31}^{\ast}V_{31}|K_{cb}|$&$\frac{g^2|K_{cb}|(m_s^2u_2+m_b^2u_3)}
{2M_W^2 \cos^2 \beta} U_{32}^{\ast}U_{32}$&$4.3 \times
10^{-4}$&$b \rightarrow se^+e^-$\\
$g^2(u_2+u_3) V_{41}^{\ast}V_{41}|K_{cb}|
$&$\frac{g^2|K_{cb}|(m_s^2u_2+m_b^2u_3)}
{2M_W^2 \cos^2 \beta} U_{42}^{\ast}U_{42} $&$4.4 \times
10^{-4}$
&$b \rightarrow s\mu^+ \mu^-$\\
$g^2(u_2+u_3) (V_{31}^{\ast}V_{41} + V_{41}^{\ast}V_{31})|K_{cb}| $&
$\frac{g^2|K_{cb}|(m_s^2u_2+m_b^2u_3)}{2M_W^2 \cos^2 \beta} $&$2.7
\times 10^{-4}$&$b \rightarrow se^{\pm} \mu^{\mp}$\\
&$\times ( U_{32}^{\ast}U_{42}+ U_{42}^{\ast}U_{32})$&&\\
&&&\\
\hline
\end{tabular}
\end{center}

\vspace*{0.5cm}

One obtains similar constraints from semileptonic decays of the K
meson, $K \rightarrow \pi l^+l^-$ and $K
\rightarrow \pi \nu {\bar \nu}$. The experimental data on these
decays is \cite{Partdata}:
\bea
BR(K \rightarrow \pi e^+e^-) &=& (2.88 {\pm} 0.13) \times 10^{-7},\nonumber \\
BR(K \rightarrow \pi \mu^+\mu^-) &= & (7.6 {\pm} 2.1) \times
10^{-7},\nonumber \\
BR(K \rightarrow \pi \mu^+e^-) &<& 2.1 \times 10^{-10}, \nonumber \\
BR(K \rightarrow \pi e^+\mu^-) &=& 7 \times 10^{-9}.
\eea
With spontaneous R-parity violation, these decays can proceed at
tree-level through either a
${\tilde u}$ or
${\tilde c}$ exchange, giving the constraints in Table 2.

The decay $K \rightarrow \pi \nu {\bar \nu}$ has an extremely small
branching ratio \cite{Partdata}:
\bea
BR(K \rightarrow \pi \nu {\bar \nu})= (1.5^{+3.4}_{-1.2}) \times 10^{-10}
\eea
These decay can proceed at tree-level through either the exchange of
a $\tilde d$ or $\tilde s$, with the bounds in Table 2.

Finally, we look at the semileptonic decays of the D-meson, $D
\rightarrow {\bar K}^{0 \ast}e^+ \nu_e$ and  $D
\rightarrow {\bar K}^{0 \ast}\mu^+ \nu_{\mu}$. The experimental
inputs used are \cite{Partdata}:
\bea
BR(D \rightarrow {\bar K}^{0 \ast}e^+ \nu_e) & = & (4.8 \pm 0.5)
\times 10^{-2} \nonumber \\
BR(D \rightarrow {\bar K}^{0 \ast}e^+ \nu_e) & = & (4.4 \pm 0.6) \times 10^{-2}
\eea
These decays can occur at tree-level though the exchange of either a
$\tilde s$ or $\tilde c$ and we derive the constraints given in Table 2.

\begin{center}
{{\bf Table 2}: Analytic bounds on mixing matrices of chargino-leptons
$U_{ij},~V_{ij}$ (with $j=3,4,5$
corresponding to $e,~
\mu$ and $\tau$) and neutralino-neutrinos $N_{ij}$ (with $j=5,6,7$
corresponding to $\nu_e,
\nu_{\mu}$ and $\nu_{\tau}$) from K and D rare semileptonic
decays.}
\vskip0.2in
\begin{tabular}{|l|l|l|c|}
\hline
gaugino type &
higgsino type
& bound & process
\\
\hline
$g^2
V_{31}V_{31}^{\ast}|K_{us}|(u_1+u_2)$&$\frac{g^2|K_{us}|(m_u^2u_1+m_c^2u_2)}
{2M_W^2 \cos^2
\beta} U_{32}^{\ast}U_{32}$&$1.4 \times 10^{-4}$&
$K^+ \rightarrow \pi^+e^+e^-$\\
$g^2 V_{41}V_{41}^{\ast}|K_{us}|(u_1+u_2)$&
$\frac{g^2|K_{us}|(m_u^2u_1+m_c^2u_2)}{2M_W^2 \cos^2 \beta}
U_{42}^{\ast}U_{42}$&$1.4 \times 10^{-4}$&
$K^+ \rightarrow \pi^+\mu^+\mu^-$\\
$g^2 (V_{31}V_{41}^{\ast}+V_{31}^{\ast}V_{41})|K_{us}|(u_1+u_2)$&
$\frac{g^2|K_{us}|(m_u^2u_1+m_c^2u_2)}{2M_W^2 \cos ^2
\beta}$&
$1.4 \times 10^{-4}$& $K^+ \rightarrow \pi^+(e^+\mu^-$\\
&$\times (U_{32}^{\ast}U_{42}+U_{42}^{\ast}U_{32})$&&$+\mu^+ e^-)$\\
&&&\\
$ [ee_dN_{51}-\frac {g(1/2+e_d\sin^2
\theta_W)}{\cos \theta_W}N_{52}]^2$
&$\frac{g^2|K_{us}|(m_d^2d_1+m_s^2d_2)}{2M_W^2 \cos ^2
\beta}N_{53}^{\ast}N_{53}$
&$1.6 \times 10^{-5}$&$K \rightarrow \pi \nu {\bar \nu}$\\
$\times 2|K_{us}|(d_1+d_2)$&&&\\
&&&\\
$-\sqrt{2} (V_{31}[ee_dN_{51}-\frac {g(1/2+e_d\sin^2
\theta_W)}{\cos \theta_W}N_{52}]d_2$&$\frac{g^2 m_c m_d}{2M_W^2 \cos
\beta \sin \beta}$&$0.13$&$D \rightarrow {\bar K}^{0 \ast}e^+ \nu_e$\\
$+V_{31}[ee_uN_{51}+\frac {g(1/2-e_d\sin^2
\theta_W)}{\cos \theta_W}N_{52}]u_2)$&$\times
(V_{32}^{\ast}N_{53} d_2+U_{32}N_{54}^{\ast} u_2)$&&\\
&&&\\
$-\sqrt{2} (V_{41}[ee_dN_{51}-\frac {g(1/2+e_d\sin^2
\theta_W)}{\cos \theta_W}N_{62}]d_2$&
$\frac{g^2 m_c m_d}{2M_W^2 \cos \beta \sin
\beta}$&0.09&$D \rightarrow {\bar K}^{0 \ast}\mu^+ \nu_{\mu}$\\
$+ V_{41}[ee_uN_{61}+\frac {g(1/2-e_d\sin^2
\theta_W)}{\cos \theta_W}N_{62}]u_2)$&
$\times (V_{42}^{\ast}N_{63} d_2+U_{42}N_{64}^{\ast} u_2)$&&\\
\hline
\end{tabular}
\end{center}

\vspace*{0.5cm}

\subsection{\bf Rare leptonic decays}

We now turn to the analysis of decays of the $\tau$ or $\mu$.
     We denote the three-body leptonic decays of the $\mu$ or $\tau$
     by $l_i \rightarrow l_j l_k l_m$, where $i, j, k, m$ are generation
indeces. The experimental bounds on
these lepton flavor violating decays are \cite{Partdata}:
\bea
BR(\mu^- \rightarrow e^-e^-e^+) &<&1 \times 10^{-12}, \nonumber \\
BR(\tau^- \rightarrow e^+\mu^-\mu^-) &<& 2.9 \times 10^{-6}, \nonumber \\
BR(\tau^- \rightarrow e^-e^-e^+) &<& 1.5 \times 10^{-6}, \nonumber \\
BR(\tau^- \rightarrow e^-\mu^-\mu^+) &<& 1.8 \times 10^{-6}, \nonumber  \\
BR(\tau^- \rightarrow e^-e^-\mu^+) &<& 1.5 \times 10^{-6}, \nonumber \\
BR(\tau^- \rightarrow \mu^-\mu^-\mu^+) &<& 1.9 \times 10^{-6}, \nonumber \\
BR(\tau^- \rightarrow \mu^-e^-e^+) &<& 1.7 \times 10^{-6}.
\eea
These decays proceed by
an exchange of a sneutrino ${\tilde \nu}_i$.
We have tabulated the constraints, both for the gaugino- and
higgsino-type couplings in Table 3.

We turn next to $\mu-e$ conversion. The conversion in nuclei is one
of the most restricted leptonic
phenomena. The upper limits extracted at PSI by the SINDRUM II
experiments are \cite{Ti,Pb}:
\bea
R_{\mu e^-} & < &  6.1 \times 10^{-13}~~~ \mbox{for}~ ^{48}Ti~ \mbox{target}\\
R_{\mu e^-} & < & 4.6 \times 10^{-11}~~~\mbox{for}~ ^{208}Pb~ \mbox{target}
\eea
This process is governed by the following effective Lagrangian:
\bea
{\cal L}_{eff}=\frac12 {\bar e}_L \gamma_{\alpha} \mu_L \left
[A^d_{\mu Ti} {\bar d}_R
\gamma_{\alpha}d_R + A^u_{\mu Ti} {\bar u}_R
\gamma_{\alpha}u_R \right ]+ \frac12 \left [S^{d,1}_{\mu Ti} {\bar
e}_L \mu_R {\bar d}_R d_L+
S^{d,2}_{\mu Ti} {\bar e}_R \mu_L {\bar d}_L d_R \right ]
\eea
It can occur at tree-level through ${\tilde d}$ or ${\tilde
\nu}$ quark exchange and it provides one of
the most stringent bounds on mixing elements, as seen in Table 3.

The effective Lagrangian for muonium $M-{\bar M}$ conversion has a
   $(V-A) \times (V-A)$ structure as in the
original papers \cite{fw}:
\begin{equation}
{\cal H}=G_{M{\bar M}} {\bar \psi}_{\mu} \gamma_{\lambda}(1-\gamma_5)
\psi_e {\bar \psi}_{\mu}
\gamma_{\lambda}(1- \gamma_5) \psi_e
\end{equation}
     where the constant $G_{M{\bar M}}$ contains information on physics
beyond the Standard Model.
In our case, the process $\mu^+ e^-
\rightarrow \mu^- e^+$, forbidden in the Standard Model, can proceed
through tree-level graphs with either
$\tilde \nu$ or $\tilde d$ exchanges  (bounds obtained are in Table
3).

     Next we investigate constraints coming from lepton family number
violating decays of tau into a meson and a
lepton, $\tau \rightarrow l+PS$ (or $V$), where $l=e$ or $\mu$,
$PS=\pi^0, \eta$, or $K^0$, and $V=\rho^0,
\omega, K^{\ast}$, or $\phi$.  The amplitude obtained from the
effective Lagrangians is:
\bea
{\cal M}(\tau \rightarrow l_k+V)& = &\frac{1}{8} A_V f_V m_V
\epsilon_{\mu}^{\ast} {\bar l}_k
\gamma^{\mu}(1-\gamma_5)\tau \nonumber \\
{\cal M}(\tau \rightarrow l_k+PS)& = & {\bar l}_k (A_L^{PS}
P_L+A_R^{PS} P_R) \tau
\eea
The bounds on the gaugino and higgsino couplings come from the
experimental data on the corresponding
decays \cite{Partdata}:
\bea
BR(\tau^- \rightarrow e^-\pi^0) &<& 3.7 \times 10^{-6}, \nonumber \\
BR(\tau^- \rightarrow \mu^-\pi^0) &<& 4.0 \times 10^{-6}, \nonumber \\
BR(\tau^- \rightarrow e^- K^0) &<& 1.3 \times 10^{-3}, \nonumber \\
BR(\tau^- \rightarrow \mu^- K^0) &<& 1.0 \times 10^{-3}, \nonumber \\
BR(\tau^- \rightarrow e^-\eta) &<& 8.2 \times 10^{-6}, \nonumber \\
BR(\tau^- \rightarrow \mu^-\eta) &<& 9.6 \times 10^{-6}, \nonumber \\
BR(\tau^- \rightarrow e^-\rho^0) &<& 2.0 \times 10^{-6}, \nonumber \\
BR(\tau^- \rightarrow \mu^-\rho^0) &<& 6.3 \times 10^{-6}, \nonumber \\
BR(\tau^- \rightarrow e^- K^{0 \ast}) &<& 5.1 \times 10^{-6}, \nonumber \\
BR(\tau^- \rightarrow \mu^- K^{0 \ast}) &<& 7.4 \times 10^{-6}.
\eea
Both of these types of decays occur at tree-level through a $\tilde
u$ or a $\tilde d$
exchange.
As seen in Table 3, the constraints from $\tau \rightarrow l_k\phi$
are too weak to give any significant bounds on R-violating couplings.

\newpage

\begin{center}
{{\bf Table 3}: Analytic bounds on mixing matrices of chargino-leptons
$U_{ij},~V_{ij}$ (with $j=3,4,5$
corresponding to $e,~
\mu$ and $\tau$) and neutralino-neutrinos $N_{ij}$ (with $j=5,6,7$
corresponding to $\nu_e,
\nu_{\mu}$ and $\nu_{\tau}$) from rare leptonic
decays.}
\vskip0.2in
\begin{tabular}{|l|l|l|c|}
\hline
gaugino type &
higgsino type
& bound & process
\\
\hline
$g^2V_{41}^{\ast}V_{31} n_1$ & $\frac{g^2 m_e^2}{2 M_W^2 \cos^2 \beta}
U_{42}^{\ast}V_{32} n_1 $& $6.6 \times 10^{-7}$ &
$\mu \rightarrow 3 e $ \\
$g^2V_{51}^{\ast}V_{41} n_2$ & $\frac{g^2 m_{\mu}^2}{2 M_W^2 \cos^2
\beta} U_{52}^{\ast}V_{42} n_2$ & $6.4 \times 10^{-3}$& $\tau
\rightarrow 3 \mu$ \\
$g^2V_{51}^{\ast}V_{31} n_1$ & $\frac{g^2 m_e^2}{2 M_W^2 \cos^2 \beta}
U_{52}^{\ast}V_{32} n_1$ & $5.6 \times 10^{-3}$ & $\tau
\rightarrow 3 e$\\
$g^2V_{51}^{\ast}V_{41} n_1$ & $\frac{g^2 m_e^2}{2 M_W^2 \cos^2 \beta}
U_{52}^{\ast}V_{42} n_1$ & $5.7 \times 10^{-3}$ & $\tau
\rightarrow 2e \mu$  \\
$g^2V_{51}^{\ast}V_{41} n_2$ & $\frac{g^2 m_{\mu}^2}{2 M_W^2 \cos^2
\beta} V_{51}^{\ast}V_{42} n_2$ & $6.2 \times 10^{-3}$& $\tau
\rightarrow 2 \mu e$\\
$g^2 V_{41}^{\ast} V_{31}d_1$ & $\frac{g^2 m_u^2}{2 M_W^2 \sin^2
\beta} V_{42}V_{32}^{\ast} n_1$ & $6.2 \times 10^{-7}$ & $\mu - e$\\
$g^2 (V_{41}^{\ast} V_{41} n_1+ V_{31}^{\ast} V_{31} n_2)$&
$\frac{g^2 }{2 M_W^2 \cos^2 \beta} (m_3^2 U_{42}^{\ast}U_{42}n_1+
m_{\mu}^2 U_{32}^{\ast}U_{32} n_2)$ & $6.3 \times 10^{-3}$ &
$M-\bar M$ \\
$g^2 V_{51}^{\ast}V_{31}(d_1+u_1)$&$\frac{g^2}{2M_W^2\sin^2 \beta}
V_{52}^{\ast}V_{32}m_u^2d_1+
\frac{g^2}{2M_W^2\cos^2 \beta}V_{52}^{\ast}V_{32}m_d^2u_1$ &
$3.5 \times 10^{-3}$ & $\tau \rightarrow e \rho$\\
$g^2 V_{51}^{\ast}V_{31}|K_{us}|(u_1+u_2)$&$\frac{g^2}{2M_W^2\cos^2
\beta}V_{52}^{\ast}V_{32}|K_{us}|(m_d^2u_1+m_s^2 u_2)$&$3.0 \times
10^{-3}$ & $\tau \rightarrow e K^{0 \ast}$\\
$g^2 V_{51}^{\ast}V_{41}(d_1+u_1)$&$\frac{g^2}{2M_W^2\sin^2 \beta}
V_{52}^{\ast}V_{42}m_u^2d_1+
\frac{g^2}{2M_W^2\cos^2 \beta}V_{52}^{\ast}V_{42}m_d^2u_1$&$4.2 \times
10^{-3}$&$\tau \rightarrow \mu \rho$\\
$g^2 V_{51}^{\ast}V_{41}|K_{us}|(u_1+u_2)$ & $\frac{g^2}{2M_W^2\cos^2
\beta}V_{52}^{\ast}V_{42}|K_{us}|(m_d^2u_1+m_s^2 u_2) $ & $3.8 \times
10^{-3}$& $ \tau \rightarrow \mu K^{0 \ast}$\\
$g^2 V_{51}^{\ast}V_{31}(d_1+u_1)$ & $\frac{g^2}{2M_W^2\sin^2 \beta}
V_{52}^{\ast}V_{32}m_u^2d_1-
\frac{g^2}{2M_W^2\cos^2 \beta}V_{52}^{\ast}V_{32}m_d^2u_1$&
$6.6 \times 10^{-2}$ & $\tau \rightarrow e \pi^0$\\
$g^2 V_{51}^{\ast}V_{31}|K_{us}|(u_1+u_2)$ & $\frac{g^2}{2M_W^2\cos^2
\beta}V_{52}^{\ast}V_{32}|K_{us}|(m_d^2u_1+m_s^2 u_2)$ &
$4.0 \times 10^{-1}$ & $\tau \rightarrow e K^{0}$\\
$g^2 V_{51}^{\ast}V_{41}(d_1+u_1)$ & $\frac{g^2}{2M_W^2\sin^2 \beta}
V_{52}^{\ast}V_{42}m_u^2d_1-
\frac{g^2}{2M_W^2\cos^2 \beta}V_{52}^{\ast}V_{42}m_d^2u_1$ &
$3.7 \times 10^{-2}$ & $\tau \rightarrow \mu \pi^0$\\
$g^2 V_{51}^{\ast}V_{31}(d_1+u_1-2u_2)$ & $\frac{g^2}{2M_W^2\sin^2
\beta}
V_{52}^{\ast}V_{42}m_u^2d_1$ & $7.8 \times 10^{-2}$&
$\tau \rightarrow e \eta$\\
&$+
\frac{g^2}{2M_W^2\cos^2 \beta}V_{52}^{\ast}V_{42}(m_d^2u_1-2m_s^2
u_2)$ &&\\
$g^2 V_{51}^{\ast}V_{41}|K_{us}|(u_1+u_2)$ & $\frac{g^2}{2M_W^2\cos^2
\beta}V_{52}^{\ast}V_{42}|K_{us}|(m_d^2u_1+m_s^2 u_2)$ &
$3.6 \times 10^{-1}$ &$\tau \rightarrow \mu K^{0}$\\
$g^2 V_{51}^{\ast}V_{41}(d_1+u_1-2u_2)$ & $\frac{g^2}{2M_W^2\sin^2
\beta}V_{52}^{\ast}V_{42}m_u^2d_1$
& $ 8.2 \times 10^{-2} $ & $\tau \rightarrow \mu \eta$\\
&$+
\frac{g^2}{2M_W^2\cos^2 \beta}V_{52}^{\ast}V_{42}(m_d^2u_1-2m_s^2u_2)$&&\\
\hline
\end{tabular}
\end{center}

\vspace*{0.5cm}

\newpage

     In the Table 4 below, we summarize our restrictions on R-violating
mixing matrix elements from tree-level processes for a set of values
of soft masses and $\tan\beta$.

\begin{center}
{{\bf Table 4}: Numerical bounds on mixing matrices of chargino-leptons
$U_{ij},~V_{ij}$ and neutralino-neutrinos $N_{ij}$ from rare decays for
$m_{\tilde f}=100$ GeV.
Quark masses have been taken as $m_u=5$ MeV, $m_d=10$ MeV, $m_c=1.5$
GeV, $m_s=200$ MeV, and
$m_b=4.5$ GeV; and $\tan\beta=2$.}
\vskip0.2in
\begin{tabular}{|l|c|r|}
\hline
Bound &
Process
&type
\\
\hline
     $|V_{31}|<0.027,~ 0.113$ & $K \rightarrow \pi e^+e^-$, $b\rightarrow
se^+e^-$ & g
\\
     $|U_{32}|<0.234,~ 0.304$ & $K \rightarrow \pi e^+e^-$, $b \rightarrow
se^+e^-$ & h
\\
     $|V_{41}|<0.027,~ 0.113$ & $K \rightarrow \pi \mu^+\mu^-$, $b \rightarrow
s\mu^+ \mu^-$ & g
\\
$|U_{42}|<0.234,~ 0.308$ & $K \rightarrow \pi \mu^+\mu^-$, $b \rightarrow
s\mu^+ \mu^-$ & h
\\
     $Re(V_{31}^*V_{41})<0.014,~ 0.008$ & $K \rightarrow \pi \mu^{\pm}
e^{\pm}$,
$b\rightarrow s \mu^{\pm}e^ {\pm}$ &g
\\
$|U_{32}U^*_{42}|<.027,~0.029$ & $K \rightarrow \pi \mu^{\pm} e^{\pm}$,
$b\rightarrow s \mu^{\pm}e^ {\pm}$ &h
\\
     $Re(V_{51}^*V_{31})<0.686$ & $B_d\rightarrow e^{\pm} \tau^{\pm}$&
g
\\
$Re(V_{51}^*V_{41})<0.876$ & $B_d\rightarrow \mu^{\pm} \tau^{\pm}$&
g
\\
$|V_{41}^*V_{31}|<1.56 \times 10^{-6},~1.67 \times 10^{-7}$ & $\mu
\rightarrow 3e,~\mu-e$ conversion & g
\\
$|V_{51}^*V_{41}|<0.015,~0.0135,~0.004,~0.02$ & $\tau \rightarrow 3
\mu,~\tau
\rightarrow 2\mu e,~\tau \rightarrow \mu \rho,~\tau \rightarrow \mu K^{0*}$
& g
\\
$|V_{51}^*V_{31}|<0.013,~0.015,~0.004$ & $\tau \rightarrow 3e,~\tau
\rightarrow
\mu 2e,~\tau \rightarrow e \rho$ & g
\\
$|0.31N_{51}+0.94N_{52}|<0.07$ & $K \rightarrow \pi \nu \nu$ &
g
\\
$|N_{53}|<0.56$ & $K \rightarrow \pi \nu \nu$ &
h
\\
$|V_{31}^*(0.93N_{51}+1.71N_{52})|<.786,~.045,~.425$ &$b\rightarrow u e
{\bar \nu},~b \rightarrow c e {\bar \nu},~D \rightarrow{\bar K}^{0*} {\bar
e}
\nu_e$ & g
\\
$|N_{53}U^*_{32}+N_{54}^*V_{32}|<0.116,~.0066$ & $b\rightarrow u e
{\bar \nu},~b \rightarrow c e {\bar \nu}$ & h
\\
$|V_{31}^*(0.93N_{61}+1.71N_{62})|<.786,~.045,$ &$b\rightarrow u e
{\bar \nu},~b \rightarrow c e {\bar \nu}$ & g
\\
$|N_{63}U^*_{32}+N_{64}^*V_{32}|<0.116,~.0066$ & $b\rightarrow u e
{\bar \nu},~b \rightarrow c e {\bar \nu}$ & h
\\
$|V_{31}^*(0.93N_{71}+1.71N_{72})|<.786,~.045$ &$b\rightarrow u e
{\bar \nu},~b \rightarrow c e {\bar \nu}$ & g
\\
$|N_{73}U^*_{32}+N_{74}^*V_{32}|<0.116,~.0066$ & $b\rightarrow u e
{\bar \nu},~b \rightarrow c e {\bar \nu}$ & h
\\
$|V_{41}^*(0.93N_{51}+1.71N_{52})|<.868,~.045$ &$b\rightarrow u \mu
{\bar \nu},~b \rightarrow c \mu {\bar \nu}$ & g
\\
$|N_{53}U^*_{42}+N_{54}^*V_{42}|<0.358~,.0066$ & $b\rightarrow u \mu
{\bar \nu},~b \rightarrow c \mu {\bar \nu}$ & h
\\
$|V_{41}^*(0.93N_{61}+1.71N_{62})|<.868,~.045,~.295$ &$b\rightarrow u \mu
{\bar \nu},~b \rightarrow c \mu {\bar \nu},~D \rightarrow{\bar K}^{0*}
{\bar
\mu}
\nu_{\mu}$ & g
\\
$|N_{63}U^*_{42}+N_{64}^*V_{42}|<0.358,~.0066$ & $b\rightarrow u \mu
{\bar \nu},~b \rightarrow c \mu {\bar \nu}$ & h
\\
$|V_{41}^*(0.93N_{71}+1.71N_{72})|<.868,~.045$ &$b\rightarrow u \mu
{\bar \nu},~b \rightarrow c \mu {\bar \nu}$ & g
\\
$|N_{73}U^*_{42}+N_{74}^*V_{42}|<0.358~,.0066$ & $b\rightarrow u \mu
{\bar \nu},~b \rightarrow c \mu {\bar \nu}$ & h
\\
$|V_{51}^*(0.93N_{51}+1.71N_{52})|<.505,~.045$ &$b\rightarrow u \tau
{\bar \nu},~b \rightarrow c \tau {\bar \nu}$ & g
\\
$|N_{53}U^*_{52}+N_{54}^*V_{52}|<0.274~,.0066$ & $b\rightarrow u \tau
{\bar \nu},~b \rightarrow c \tau {\bar \nu}$ & h
\\
$|V_{51}^*(0.93N_{61}+1.71N_{62})|<.505,~.045,$ &$b\rightarrow u \tau
{\bar \nu},~b \rightarrow c \tau {\bar \nu}$ & g
\\
$|N_{63}U^*_{52}+N_{64}^*V_{52}|<0.274,~.0066$ & $b\rightarrow u \tau
{\bar \nu},~b \rightarrow c \tau {\bar \nu}$ & h
\\
$|V_{51}^*(0.93N_{71}+1.71N_{72})|<.505,~.045$ &$b\rightarrow u \tau
{\bar \nu},~b \rightarrow c \tau {\bar \nu}$ & g
\\
$|N_{73}U^*_{52}+N_{74}^*V_{52}|<0.274,.0066$ & $b\rightarrow u \tau
{\bar \nu},~b \rightarrow c \tau {\bar \nu}$ & h
\\
\hline
\end{tabular}
\end{center}

(By "h" and "g" we mean higgsino or gaugino coupling.)

\vspace*{0.5cm}

\section{One Loop Processes}

In addition to processes that can occur at tree-level, there are
others which can only occur at one
loop-level, but are highly suppressed; or processes like $\mu-e$
conversion, which may set more stringent
limits at one-loop level than at tree-level. Usually these processes
invove chirality flip on an internal or
external leg. These are: the anomalous magnetic moment of the muon
$(g-2)_{\mu}$, lepton flavor violating
processes $\mu-e$ conversion and
$\mu \rightarrow e \gamma$ and the lepton and quark electric dipole moments.

First we investigate the effect of spontaneous R-parity breaking on
the decay $\mu \rightarrow e
\gamma$. The amplitude of the $\mu\rightarrow e \gamma$ transition
can be written in the
form of the usual dipole-type interaction:
\bea
{\cal M}_{\mu \rightarrow e \gamma}
=\frac{1}{2}\bar{\psi}_e(d_LP_L+d_RP_R)\sigma^{\mu\nu}F_{\mu\nu}\psi_\mu
\label{eq:ampl}
\eea
It leads to the branching ratio:
\bea
\label{eq:exp}
BR(\mu\rightarrow e \gamma)=\frac{1}{16\pi} \tau_{\mu}(|d|_L^2+ |d|_R^2)m^3_\mu
\eea
Comparing it with the standard decay width, $\Gamma_{\mu\rightarrow e
\nu\bar{\nu}}=\frac{1}{192\pi^3}G_F^2m_{\mu}^5$ and using the experimental
constraint on the branching ratio $B.R.(\mu \rightarrow e \gamma)
<1.2 \times 10^{-11}$ \cite{Partdata},
one obtains the following limit on  the dipole
amplitude:
\bea
|d|=\sqrt{(|d_L|^2+|d_R|^2)/2}<1.73 \cdot 10^{-26}~e\cdot cm
\label{eq:limit}
\eea

A non-vanishing dipole interaction results in a fermion chirality flip.
There are two possibilities for this to occur. One is
that the chirality flip occurs on the external muon line, resulting
in a proportionality of the decay
amplitude to the muon mass. The other is that the chirality flip occurs on the
internal line, resulting in proportionality of the same amplitude to
the mass of the
     fermion in the loop. This latter process requires the mixing of the
left and right squarks or
sleptons, and the resulting amplitude is proportional to the mixing
angle. In R-parity
conserving SUSY, the latter
process dominates due to the large fermion  mass in the loop,  and
also due to the loop function
which is larger (by an order of magnitude or more) than the
corresponding one for the process
with external chirality flip. The same is true with spontaneous
R-parity breaking
and the bound obtained is:
\bea
\frac{1}{16 \pi^2}\frac{g^2|V_{j1}U^{\ast}_{j2}|}{2 M_W \cos \beta}
m_{\mu}\frac{f_1(x)}{
m_{{\tilde f}_1}^2} < 1.47
\times 10^{-3}
\eea
where
the loop integral is:
\bea
f_1(x) &=& \frac{1}{2(1-x)^2}\left[ 3-x+\frac {2 \ln x}{1-x} \right]
\label{loopint}
\eea
and $x=\frac{m_{\tau}^2}{m_{{\tilde f}}^2}$.

$\mu -e$ conversion in nuclei is perhaps the most interesting
lepton-flavor violating process
experimentally. From a theoretical point of view, it is the most difficult to
disentangle, because of the interdependence between particle and
nuclear physics elements,
in particular the difficulty in evaluating nuclear matrix elements.
The process is very interesting at one-loop level from two points of
view. First, it has quite a different
structure from
$\mu
\rightarrow e
\gamma$  (as opposed to $\mu^+ \rightarrow e^+ e^ +e^-$). Therefore it provides
complimentary information on muon decay from the first two decays:
     it can occur even when $\mu \rightarrow e \gamma$ is forbidden, and it
could be a better indicator of a rich gauge structure, such as extra Z or W
bosons.
Second, it has been shown that for a class of models $\mu-e$
conversion is enhanced with respect to $\mu
\rightarrow e \gamma$ by large $\ln(m_{\mu}^2/\Lambda^2)$, where
$\Lambda$ is the scale responsible for the new
physics \cite{rs}. With the expected improvement in experimental
data, this test is likely to become the most
stringent in R-parity violation. Based on the above transition
elements, the branching ratio for the coherent
$\mu^- -e^-$ conversion is given by \cite{weinfein}:
\bea
R_{ph}(\mu^-N \rightarrow e^-N)&=& \frac{p_eE_e Z\alpha^5
Z_{eff}^4F_p^2}{m_{\mu}
\Gamma_{capt}}
\{|f_{E0}(-m_{\mu}^2)+f_{M1}(-m_{\mu}^2)+f_{M0}(-m_{\mu}^2)+f_{E1}(-m_{\mu}^2)
|^2
\nonumber \\
&+&|f_{E0}(-m_{\mu}^2)+f_{M1}(-m_{\mu}^2)-f_{M0}(-m_{\mu}^2)-f_{E1}
(-m_{\mu}^2)|^2 \}
\eea
where $\Gamma_{capt}$ is the total muon capture rate, $Z_{eff}$ is an
effective atomic
charge obtained by averaging the muon wave function over the nuclear
density, and $F_p$
is the nuclear matrix element. The functions  $f_{E0},f_{E1},f_{M0}$
and $f_{M1}$ depend on loop functions and
on the R-parity violating couplings \cite{rs}. The bounds obtained,
listed in Table 4, restrict the same
combination of parameters as the bounds obtained from $\mu
\rightarrow e \gamma$ but slightly stronger.
As for the radiative decays of the $\tau$ lepton, $\tau \rightarrow
\mu \gamma$ and $\tau \rightarrow e
\gamma$, they do not add anything new to the the bounds found so far.
These radiative decays constrain the same combination of mixing
matrix elements, but the bounds are much
weaker, owing to weaker experimental limits
of the radiative decays of the $\tau$ versus the $\mu$.

Next we evaluate the contributions coming from the electric dipole
moments. The electric dipole moment of an
elementary fermion is defined through its  electromagnetic form
factor $F_3(q^2)$ found from the (current)
matrix element:
\bea
\label{formfactors}
\langle f(p')|J_{\mu}(0)|f(p) \rangle=\bar{u}(p')\Gamma_{\mu}(q)u(p),
\eea
where $q=p'-p$ and
\bea
\label{current}
\Gamma_{\mu}(q)=F_1(q^2)\gamma_{\mu}+F_2(q^2)i\sigma_{\mu\nu}q^{\nu}/2m
+F_A(q^2)
(\gamma_{\mu}\gamma_5q^2-2m\gamma_5q_{\mu})+F_3(q^2)\sigma_{\mu\nu}
\gamma_5q^{\nu}/2m,
\eea
with $m$ the mass of the fermion. The EDM of the fermion field $f$ is
then given
by
\bea
\label{edm}
d_f=-F_3(0)/2m,
\eea
corresponding to the effective dipole interaction
\bea
\label{dipole}
{\cal L}_I= - \frac{i}{2} d_f \bar{f}\sigma_{\mu\nu}\gamma_5 f F^{\mu\nu}
\eea
     The effective Lagrangian is induced at one-loop level if the theory
contains a CP-violating coupling at tree-level.
     We can parametrize the interaction of a fermion
$\Psi_f$ with other fermions $\Psi_i$-s and scalars $\Phi_k$-s, with
respective charges $Q_f,~Q_i$ and $Q_k$,
in general as:
\bea
-{\cal L}_{int} = \sum_{ik} {\bar \Psi}_f\left( A_{ik}
\frac{1-\gamma_5}{2} + B_{ik}
\frac{1+\gamma_5}{2}\right) \Psi_i \Phi_k +H.C.
\eea
If there is CP-violation, then $Im(A_{ik}B^*_{ik}) \neq 0$, and the
one-loop fermion EDM
is given by:
\bea
d^E_f=\sum_{ik} \frac{m_i}{(4\pi)^2m_k^2}Im(A_{ik}B^*_{ik})\left ( Q_i
f_1(\frac{m_i^2}{m_k^2})+Q_k f_2(\frac{m_i^2}{m_k^2})\right )
\eea
with:
\bea
f_2(x)= \frac{1}{2(1-x)^2}\left ( 1+x +\frac{2 x \ln x}{1-x}\right ) ,
\eea
assuming charge conservation at the vertices $Q_k=Q_f-Q_i$. Since a
non-vanishing $d_f$
in the SM results in fermion chirality flip, it  requires both
$CP$ violation and $SU(2)_L$ symmetry breaking.
Experimentally, the EDMs of the electron and the neutron are some of the most
restrictive parameters in the Particle Data Booklet, the present
experimental upper
limits being
$d_e \leq 4.3 \cdot 10^{-27} ecm$ and $d_n \leq 6.3 \cdot 10^{-26}
ecm$ \cite{Partdata}.

The spontaneous R-violating contribution to the dipole moment of an
electron is:
\bea
d_{e}^E= \frac{\alpha_{EM}}{4 \pi\sin^2 \theta_W}
Im(V_{i1}U_{i2}) \frac {m_e m_{e_i}}{\sqrt {2} M_W \cos \beta} \frac{f_1(x_e)}
{m_{{\tilde f}}^2} < 4.3 \times 10^{-27} ecm
\eea
with $x_e=\frac{m_{e_i}^2}{m_{{\tilde f}}^2}$ and the d-quark contribution is:
\bea
&d_{d}^E&= N_c\frac{\alpha_{EM}}{4 \pi\sin^2 \theta_W}
\left \{Im(V_{i1}U_{i2}) \frac {m_d m_{e_i}}{\sqrt {2} M_W \cos \beta}
\frac{f_1(x_i)}{
m_{{\tilde f}}^2} +\sqrt {2}\left [ e e_d Im(N_{71}N_{73})
\right.\right. \nonumber \\
&-&\left. \left.\frac{g}{\cos \theta_W}(-\frac{1}{2}-e_d \sin^2
\theta_W)Im(N_{72}N_{73})\right] \frac{F(x_{\nu})}{
m_{{\tilde d}}^2} \right \} < 4.725 \times 10^{-26} ecm
\eea
with $F=f_1+2f_2$, $x_{\nu}=\frac{m_{\nu}^2}{m_{{\tilde f}}^2}$.
To evaluate the EDM of the neutron we use:
\bea
d_n=\frac43 d_d-\frac13 d_u
\eea

We include for completeness the constraint arising from the new
measurement of the anomalous magnetic moment of
the muon. The new measurement for the muon anomalous magnetic moment
$a_{\mu}$ corresponds
to a deviation from the Standard Model prediction:
\begin{equation}
a_{\mu}^{exp}-a_{\mu}^{SM} = (4.26 \pm 1.65) \times 10^{-9}
\end{equation}
If the deviation can be attributed to new physics effects, then at
90\% C.L. $\delta a_{\mu}^{NP}$
must lie in the range:
\begin{equation}
2.15 \times 10^{-9} \leq \delta a_{\mu}^{NP} \leq 6.37 \times 10^{-9}
\end{equation}
The anomalous magnetic moment of the muon arises from terms of the form:
\bea
\frac{ie}{2m_{\mu}} F(q^2){\bar \psi} \sigma_{\alpha \beta}q^{\beta} \psi
\eea
with $a_{\mu} =F(0)$.
The contributions to $a_{\mu}$ are
proportional to the mass of the muon squared:
\bea
\delta a_{\mu}=\frac{m_{\mu}^2}{2}(A_L^{22}+A_R^{22})
\eea
   From spontaneous R-parity violation, we obtain the bound:
\bea
\frac{1}{4 \pi^2} \frac{g^2 m_{\mu}^2}{2M_W \cos \beta} N_{i3}[\sin
\theta_W N_{i1}-\frac{g}{\cos
\theta_W}(\frac{1}{2} -\sin^2 \theta_W)N_{i2}]\frac{f_2(x_{\nu})}{
m_{{\tilde f}}}
< 4.2 \times 10^{-9}
\eea
where $i=5,6$ or $7$. We take $m_{\nu_\tau}=1$ eV.

In Table 4 below we summarize all one-loop bounds we obtained. In the
case in which
more than a term is present in a constraint, and we have insufficient
information to bound the terms separately,
we obtain the bounds by assuming that only one term is non-zero.
All of these bounds include products of couplings from vertices
including higgsino or gaugino, and are thus new bounds, not present
in models with explicit trilinear R-parity violation.

\newpage

\begin{center}
{{\bf Table 4}: Numerical bounds on mixing matrices of chargino-leptons
$U_{ij},~V_{ij}$ and neutralino-neutrinos $N_{ij}$ from one-loop processes for
$m_{\tilde f}=100$ GeV and $\tan \beta =2$.}
\vskip0.2in
\begin{tabular}{|l|c|c|}
\hline
Bound &
Process
&type
\\
\hline
     $Re(V^*_{31}U^*_{32})< 2 \times 10^{-5}$ & $\mu \rightarrow e 
\gamma$ & g, h
\\
     $Re(V^*_{41}U^*_{42})< 3.8 \times 10^{-5}$ & $\mu \rightarrow e
\gamma$ & g, h
\\
     $Re(V^*_{51}U^*_{52})< 6.8 \times 10^{-5}$ & $\mu \rightarrow e
\gamma$ & g, h
\\
     $Re(N_{i1}N_{i3})< 1.4 \times 10^{-2}$ & $\mu \rightarrow e \gamma$ & g, h
\\
     $Re(N_{i2}N_{i3})< 8.5 \times 10^{-2}$ & $\mu \rightarrow e \gamma$ & g, h
\\
     $Re(V^*_{31}U^*_{32})< 3 \times 10^{-3}$ & $(g-2)_{\mu}$ & g, h
\\
     $Re(V^*_{41}U^*_{42})< 6.1 \times 10^{-3}$ & $(g-2)_{\mu}$ & g, h
\\
     $Re(V^*_{51}U^*_{52})< 1.1 \times 10^{-2}$ & $(g-2)_{\mu}$ & g, h
\\
$Re(V_{31}^*U_{32}^*)<  5.6 \times 10^{-6}$ & $\mu-e$ conversion & g, h
\\
$Re(V_{41}^*U_{42}^*)<  1.1 \times 10^{-5}$ & $\mu-e$ conversion & g, h
\\
$Re(V_{51}^*U_{52}^*)<  2 \times 10^{-5}$ & $\mu-e$ conversion & g, h
\\
     $Re(N_{i1}N_{i3})< 9.3 \times 10^{-3}$ & $\mu-e$ conversion & g, h
\\
     $Re(N_{i2}N_{i3})< 5.7 \times 10^{-3}$ & $\mu-e$ conversion & g, h
\\
$Im(V_{41}U_{42})< 6.2 \times 10^{-2}$ & $EDM_{e}$
& g, h
\\
$Im(V_{51}U_{52})< 6.9 \times 10^{-3}$ & $EDM_{e}$
& g, h
\\
$Im(V_{41}U_{42})< 2.9 \times 10^{-3}$ & $EDM_{n}$ & g, h
\\
$Im(V_{51}U_{52})< 1.7 \times 10^{-4}$ & $EDM_{n}$ & g, h
\\
$Im(N_{i1}N_{i3})< 2.8 \times 10^{-2}$ & $EDM_{n}$ & g, h
\\
$Im(N_{i2}N_{i3})< 9.3 \times 10^{-3}$ & $EDM_{n}$ & g, h
\\
\hline
\end{tabular}
\end{center}
where $i=5,6,7$.

\section{Conclusion}

Conservation of R-parity, introduced to distinguish particles from their
supersymmetric partners, is not imposed by any symmetry of the model.
Explicit R-parity violation, allowed in MSSM, may not be allowed by higher
gauge structures.
However, the R-parity may be broken
spontaneously through the Higgs mechanism.
This type of breaking is achieved
through  vevs for the sneutrino fields. It has the attractive feature
that it only breaks lepton number, thus avoiding fast proton decay. It allows
for a dynamical mechanism to break R, much like electroweak symmetry
breaking.

Spontaneous R parity breaking generates bilinear terms in the Lagrangian with
both gaugino- and higgsino-type couplings. In this work, we assumed a general
pattern of neutrino-neutralino and lepton-chargino mixing.
Although the particle
    content of a given supersymmetric model will have to be enlarged 
to allow for
    spontaneous R-parity breaking, we deal with a truncated version and 
assume an
effective MSSM  particle content.
We then set general constraints on mixing
matrix elements, valid for any supersymmetric model with spontaneous R parity
    breaking.
For tree-level processes, we obtain some mass-dependent bounds
and also
some mass-independent bounds which come from gaugino-type couplings,
most of which
are new.
Restricting processes which require chirality flip (at one-loop level), we
obtain strong bounds on products of gaugino and higgsino couplings all of which
    are new.
These results are complementary to those previously found \cite{prev}.

\section {\bf Acknowledgement}

This work was supported in part by NSERC under grant number SAP0105354
and Academy of Finland (projects no 48787 and no 163394). M. F. would
like to thank the
Helsinki Institute of Physics, where part of this work was done, for
their warm hospitality.

\bibliographystyle{plain}

\vskip0.2in

\end{document}